\documentclass[aps,prd,10pt,nofootinbib,twocolumn,eqsecnum,showpacs,showkeys,superscriptaddress,preprintnumbers,altaffilletter]{revtex4-1}
\usepackage{graphicx}
\usepackage{amsmath}
\usepackage{multirow}
\usepackage{hyperref}
\usepackage{soul}
\usepackage{graphicx}
\usepackage{dcolumn}
\usepackage{amssymb}
\usepackage{amsfonts}
\usepackage{amsbsy}
\usepackage{color}
\usepackage{rotating}
\usepackage[english]{babel}
\usepackage{multirow}
\usepackage{float}

\newcommand{\be}{\begin{equation}}
\newcommand{\ee}{\end{equation}}
\newcommand{\bea}{\begin{eqnarray}}
\newcommand{\eea}{\end{eqnarray}}
\newcommand{\beal}{\begin{aligned}}
\newcommand{\eeal}{\end{aligned}}
\newcommand{\bi}{\begin{itemize}}
\newcommand{\ei}{\end{itemize}}

\hypersetup{
    colorlinks=true,
    linkcolor=blue,
    filecolor=magenta,
    urlcolor=cyan,
    citecolor=cyan
}

\begin{document}

\title{Barrow nearly-extensive Gibbs-like entropy favoured by the full dynamical and geometrical data set in cosmology} 

\author{T. Denkiewicz}
\email{tomasz.denkiewicz@usz.edu.pl}
\affiliation{Institute of Physics, University of Szczecin, Wielkopolska 15, 70-451 Szczecin, Poland}
\author{V. Salzano}
\email{vincenzo.salzano@usz.edu.pl}
\affiliation{Institute of Physics, University of Szczecin, Wielkopolska 15, 70-451 Szczecin, Poland}
\author{Mariusz P. D\c{a}browski}
\email{mariusz.dabrowski@usz.edu.pl}
\affiliation{Institute of Physics, University of Szczecin, Wielkopolska 15, 70-451 Szczecin, Poland}
\affiliation{National Centre for Nuclear Research, Andrzeja So{\l}tana 7, 05-400 Otwock, Poland}
\affiliation{Copernicus Center for Interdisciplinary Studies, Szczepa\'nska 1/5, 31-011 Krak\'ow, Poland}

\date{\today}

\begin{abstract}
We apply the full set of most update dynamical and geometrical data in cosmology to the nonextensive Barrow entropic holographic dark energy. We show that the data point towards an extensive Gibbs-like entropic behaviour for the cosmological horizons, which is the extreme case of the Barrow entropy, with the entropy parameter being $\Delta > 0.86$, close to the maximum threshold of $\Delta =1$ where the fractal dimension of the area-horizon becomes almost or just the volume and the intensivity is recovered. Futhermore, we find that the standard Bekenstein area-entropy limit ($\Delta = 0$) is excluded by the set of our data. This contradicts the bounds obtained recently from early universe tests such as the baryon asymmetry, the big-bang nucleosynthesis, and the inflation limiting $\Delta< 0.008$ at the most extreme case.
\end{abstract}

\maketitle

\section{Introduction}

In view of the recent tensions in cosmology (see \cite{Abdalla:2022yfr} for a review), various non-standard approaches have been attempted as possible resolutions of these issues of the cosmological setting. One of such approaches is based on the application of the cosmological horizons, which and are rooted in the studies of the black hole event horizons and their thermodynamical framework. After the seminal works on that topic some decades ago, it is commonly known that black hole systems behave similarly to thermodynamical systems. This is generally expressed by the black hole thermodynamical laws operating the notions of the Bekenstein entropy \cite{Bekenstein:1973ur} and the Hawking temperature \cite{Hawking:1975vcx}. The former is commonly called the ``area entropy'' due to the fact that it is proportional to the area of the horizon of a black hole and, in consequence, to its radius squared. However, this poses a problem which is not so much appreciated by the scientific community: the proportionality of the entropy to the surface area makes the Bekenstein entropy nonextensive, while extensivity means that the entropy should be proportional to the volume rather than the area, as it is the case in the standard Gibbs thermodynamics. Besides, the Bekenstein entropy is also nonadditive \cite{BiroVan2011,Alonso-Serrano:2020hpb} which again differs it from the Gibbs entropy. In view of that, all applications of the cosmological horizons towards resolution of the current problems should take this into account. 

In this paper we focus ourselves on these issues while applying the full set of dynamical and geometrical data nowadays available in cosmology. The work is then organized as follows: in Sec. (\ref{sec:nonextensive}) we discuss the nature of nonextensive horizon entropies (Bekenstein, Barrow, Tsallis-Cirto models) and exhibit the relations between them; in Sec.~(\ref{sec:Model}) we introduce the main features of Barrow holographic cosmology; in Sec.~(\ref{sec:Data}) we describe the data we have used for our analysis; finally, in Sec.~(\ref{sec:Results}) we outline the main outcomes of our work.

\section{Nonextensive entropies and holographic screens in cosmology} 
\label{sec:nonextensive}

So far, there is no convincing statistical theory of black hole thermodynamics within the framework of Bekenstein and Hawking which would be based on the proper definitions of microscopic degrees of freedom resulting after appropriate averaging to phenomenological quantities. This would require a generalization of the Gibbs entropy definition to nonextensive and nonadditive cases. Generalizations of such a type have been proposed, and among one of the earliest of them, there was the proposal of Tsallis \cite{Tsallis:1987eu,tsallisbook} in which a new parameter $q$ is measuring the deviation from extensivity, with the entropy reading as 
\be
S_{T} =- k_B \sum_{i}[p(i)]^q\ln_qp(i), 
\label{S_T}
\ee 
where $p(i)$ is the probability distribution defined on a set of microstates $\Omega$, $q \in {\cal R}$ is the nonextensivity parameter, and $k_B$ is the Boltzmann constant. This tricky idea comes from the introduction of the $q$-logarithmic function $\ln_qp$  defined as
\be
\ln_q p=\frac{p^{1-q}-1}{1-q},
\ee 
such that, in the limit, $q \to 1$, Tsallis entropy (\ref{S_T}) {\it reduces to Gibbs entropy}
\be
S_G=-k_B \sum_ip(i)\ln p(i).
 \label{SG}
\ee
Tsallis entropy (\ref{S_T}) satisfies a {\it nonadditive} composition rule of the following form
\begin{eqnarray}
S_{12} = S_{1}+S_{2}+\frac{1-q}{k_B} S_{1} S_{2} ,
\label{tsalliscomp}
\end{eqnarray}
while the Bekenstein entropy fulfils another rule of nonadditivity which reads \cite{BiroVan2011,Alonso-Serrano:2020hpb} 
\be
S_{12} = S_1 + S_2 + 2 \sqrt{S_1} \sqrt{S_2} .
\ee
In fact, via the introduction of the ``formal logarithm" approach \cite{BiroVan2011}, one is able to transfer it into a corresponding additive entropy in such a way that 
\be
S_R=\frac{k_B}{1-q}\left[\ln{\left(1+\frac{1-q}{k_B} S_T \right)}\right],
\label{ERenyi}
\ee
which is the R\'enyi entropy \cite{Renyi1959}
\be
S_R= k_B \frac{\ln \sum_i p^q(i)}{1-q}. 
\label{S_R} 
\ee
As it was already mentioned, the Bekenstein entropy is nonextensive and nonadditive. However, it is often put as Tsallis entropy into (\ref{ERenyi}) in order to make it additive, though still keeping it nonextensive. The Tsallis entropy, on a similar footing as the Bekenstein entropy, can be applied to black holes. In fact, it is known as the Tsallis-Cirto entropy  \cite{Tsallis:2012js,Tsallis:2019giw} and reads 
\be
S_{\delta} = k_B  \left(\frac{S_{B}}{k_B} \right)^\delta   , 
\label{TC}
\ee
with $\delta >0$ a real parameter, and $S_B$ the Bekenstein area entropy 
\be
S_B = \frac{k_B c^3 A}{4 G \hbar}, 
 \label{BH}
\end{equation}
where $\hbar$, $G$, and $c$ are the reduced Planck constant, the Newton gravitational constant, and the speed of light, respectively. In (\ref{BH}), the main variable characteristic of the entropy is the area $A$ of the horizon. For the simplest Schwarzschild black hole case, it is defined as $A=4\pi r_h^2$, where $r_h=2GM/c^2$ is the Schwarzschild radius and $M$ is the mass of a black hole. Tsallis-Cirto entropy fulfils the composition rule in the form
\begin{equation}
    S_{12}=k_B\left[\left (\frac{S_{1}}{k_B}\right)^{1/\delta}+\left (\frac{S_{2}}{k_B}\right)^{1/\delta}\right]^\delta. \label{tsallisdeltacomp}
\end{equation}
It is worth noticing that for $\delta=3/2$, the Tsallis-Cirto entropy scales with the volume, and so it {\it becomes an extensive quantity}. 
For any other value of $\delta$, as it is clear from (\ref{TC}), this entropy scales proportionally to an area, similarly to the Bekenstein entropy. The only difference is that it introduces a new parameter $\delta$, which makes this scaling nonlinear. 

The application of the event horizon to black holes for the nonextensive entropies has led cosmologists to use them also for the cosmological horizons in order to try to solve the current problems of cosmology such as the dark energy problem or the Hubble tension problem. This is why one observes recently many applications of the nonextensive entropies to the holographic screens in the Universe. Apart from Bekenstein entropy,  also the Tsallis entropy was considered theoretically \cite{Luciano:2022viz,Luciano:2022ely} and tested cosmologically \cite{Mamon2020,AlMamon:2020usb,Asghari:2021lzu,Luciano:2022ely,Ghoshal:2021ief}. Other entropies, such as Kaniadakis entropy \cite{Kaniadakis:2002zz,Kaniadakis:2005zk,Drepanou:2021jiv,Almada2022_Kaniad_test}, have also been tested. 

Our main concern in this paper is the application of yet another case of nonextensive entropy, which is the Barrow entropy, in the cosmological setting.

Barrow entropy \cite{Barrow:2020tzx} is a proposal which, like the Bekenstein entropy, has no statistical roots, but its fractal nature possesses a clear physical interpretation - it is due to quantum fluctuations of some hierarchical scales. The idea was to carry on with the tight structure of smaller and smaller spheres being attached sequentially on the top of the previous ones from the horizon to the infinity. This forms the fully fractal structure which is known as the sphereflake, and can be referred to any textbook in dynamical systems \cite{ott}. Its mathematical description suggests no appeal to physics, but on the contrary, it seems to be quite reasonable if one appeals to the quantum theory, which allows quantum fluctuations of the horizon in some analogy to what happens in the early (inflationary) universe when the quantum fluctuations on each length scales are generated. Fully satisfactory theory of such fluctuations has yet not been attempted, but it seems well motivated. The basic derivation of the Barrow entropy formula has been presented in a few papers including our previous contribution \cite{Dabrowski:2020atl}. In short, due to the fractal structure of the horizon sphere, its area is modified such that it gains some extra fractal volume which makes its dimension larger than an area, i.e. the horizon possesses the fractal dimension $2 \leq d_f = 2 + \Delta \leq 3$, and results in an effective horizon area of $r_{eff} = r_{h}^{d_f}$, where $r_h$ is the standard, non-fractal, horizon radius. As a result, the horizon area is modified, yielding the expression for the Barrow entropy as follows
\begin{eqnarray}
S_{Barrow}=k_B \left( \frac{A}{A_p}\right)^{1+\frac{\Delta}{2}}
\label{barrowarea}
\end{eqnarray}
where $A$ is the horizon area, $A_p$ is the Planck area, and $\Delta$ is the fractal dimension parameter bound to take values between 0 and 1, i.e. $0 \leq \Delta \leq 1$.  Quick inspection of (\ref{TC}) and (\ref{barrowarea}) reveals that the Tsallis-Cirto entropy and the Barrow entropy are mathematically equivalent  provided that \cite{Abreu:2020wbz}
\begin{eqnarray}
\delta \rightarrow 1+\frac{\Delta}{2} ,
\label{barrowdelta}
\end{eqnarray}
except the $\delta$ parameter is not limited besides being required to be real. In fact, they both yield the same temperatures and heat capacities as a function of black hole masses \cite{Cimdiker2023}. However, the most striking property is that they both, in the appropriate limits ($\delta=3/2$  for Tsallis-Cirto and $\Delta=1$ for Barrow), lead to the extensivity (while still keeping nonadditivity) of the entropy for the black hole and cosmological horizons. In this limit they are closest to the standard Gibbs thermodynamic formulation. 

Since its first proposal, Barrow entropy was studied theoretically \cite{refId0,Sheykhi2021P,Saridakis_nonflat,DiGennaro:2022grw,ong_running_BE,Luciano:2022hhy,Sheykhi2023be} and also tested observationally in Refs. \cite{Luciano:2022pzg,Dabrowski:2020atl,Ghoshal:2021ief,Sheykhi2021P,Luciano:2023roh}. It is interesting to note that using the Jacobson method \cite{Jacobson1995} of obtaining gravity from thermodynamics, Barrow entropy gives a general relativity-like gravity with a rescaled cosmological constant $\tilde{\Lambda} = \Lambda[(1+\Delta/2)A^{\Delta/2}]^{-1}$, while Tsallis entropy modifies the matter side of the Einstein field equations influencing the gravitational constant, $\tilde{G} = (G/\delta)(A/A_0)^{1-\delta}$, giving in the limit $\delta \to 1$ ($\Delta \to 0$) the standard $G$ \cite{DiGennaro:2022grw}.

\section{Barrow Holographic Horizons}
\label{sec:Model}

The key point to ``translate'' the modification of the horizon area proposed by \cite{Barrow:2020tzx}, and the resulting changes in the effective Bekenstein entropy, into a cosmological context is to refer to holographic dark energy (DE) approaches \cite{Wang:2016och}. In that context, DE is given by $\rho_{DE} \propto S_{eff} L^{-4}$ where in our case the effective Bekenstein entropy is $S_{eff} \propto A_{eff} \propto L^{2+\Delta}$. The distance $L$ is \textit{a horizon length}, whose definition is not set unequivocally.

Then, we can express Barrow holographic dark energy (BH) as \cite{Saridakis:2020zol} :
\begin{equation}\label{eq:BH_dens}
\rho_{BH} = \frac{3\,C^2}{8\pi G} L^{2\left(\frac{\Delta}{2}-1\right)}\, ,
\end{equation}
where $C$ is the holographic parameter with dimensions of $[\mathsf{T}]^{-1}[\mathsf{L}]^{1-\Delta/2}$ and $G$ the Newton gravitational constant. It is worth noticing that the $\Lambda$CDM limit of (\ref{eq:BH_dens}) is obtained for $\Delta = 2$ which is \textit{beyond} the Barrow fractal parameter range, though it still is within the Tsallis parameter $\delta$ ($\delta = 2$) range (cf. \ref{barrowdelta}). The choice of which horizon to be used is an open issue. In this work we will assume it to be the future event horizon \cite{Hsu:2004ri,Li:2004rb},
\begin{equation} \label{eq:event_hor}
L \equiv a\, \int^{\infty}_{t} \frac{dt'}{a} = a \int^{\infty}_{a} \frac{da'}{H(a')a'^{2}} \, ,
\end{equation}
where $a$ is the scale factor and $H(a)$ the Hubble parameter. In the literature, one can find that also other choices are considered. For example, in \cite{Pavon:2005yx} it is shown how the Hubble horizon, 
\begin{equation}
L \equiv \frac{c}{H(a)}\, ,
\end{equation}
can be assumed as a boundary, although with some caveats. Also in \cite{Dabrowski:2020atl} we have shown how a Barrow DE fluid with the Hubble horizon can be, in principle, a healthy model to be used to describe cosmological data. But we must point out that the Hubble horizon is not a ``true horizon'', in strict terms, as it can be crossed and has been crossed in the past \cite{Davis:2003ad}.

At the same time, some issues of causality violation which might derive from the choice of the future event horizon as the boundary have been raised \cite{Li:2004rb}. In order to solve this problem, an infrared cut-off for holographic DE models has been proposed in \cite{Granda:2008dk,Granda:2008tm}, which reads as
\begin{equation}
L \equiv c \left[ \alpha H^{2}(a) + \beta \dot{H}(a) \right]^{-1/2}\, ,
\end{equation}
with $\alpha$ and $\beta$ free dimensionless parameters, and the dot being a time derivative. Even so, we stress again that here we have only focused on the choice of the future event horizon, while other cases are postponed to future works.

\subsection{Kinematic quantities}

Starting from (\ref{eq:event_hor}), we proceed using the standard first Friedmann equation, 
\begin{equation} \label{eq:fried_eq}
H^{2}(a) = \frac{8 \pi G}{3} \left[\rho_{m}(a) +\rho_{r}(a) +\rho_{BH}(a)\right]\, ,
\end{equation}
where we consider the presence of matter, radiation and DE, and the standard continuity equation for both matter and radiation, namely,
\begin{equation}
\dot{\rho}_{m,r}(a) + 3 H(a) \left[\rho_{m,r}(a) + \frac{p_{m,r}(a)}{c^{2}} \right] = 0\, ,
\end{equation}
with the pressure parameterized as $p_{i} = w_{i} \rho_{i}$, with the equation of state parameter $w_{i}$ being $0$ for standard pressureless matter and $1/3$ for radiation. If we introduce the standard density parameters $\Omega_{i}(a)$, 
\begin{equation}\label{eq:dimensionless_dens}
\Omega_{m,r}(a) = \frac{H^{2}_0}{H^{2}(a)} \Omega_{m,r} a^{-3(1+w_{m,r})} \,,
\end{equation}
\begin{equation}\label{eq:dimensionless_dens_BH}
\Omega_{BH}(a) = \frac{C^2 }{H^{2}(a)}  L^{2 \left(\frac{\Delta}{2}-1\right)}\,,
\end{equation}
then the cosmological equation (\ref{eq:fried_eq}) can be rewritten as 
\begin{equation}\label{eq:fried_eq_n1}
1 = \Omega_m(a) + \Omega_r(a) + \Omega_{H}(a)\, ,
\end{equation}
so that finally the Hubble parameter can be expressed as 
\begin{equation} \label{eq:Hubble_alt}
H(a) = H_{0} \sqrt{\frac{\Omega_{m} a^{-3} + \Omega_{r} a^{-4}}{1-\Omega_{BH}(a)}}\, .
\end{equation}
This latest expression is quite useful: if we use it into (\ref{eq:event_hor}), equate the obtained horizon length expression with the one which can be derived from inversion of (\ref{eq:dimensionless_dens_BH}), and  differentiate both the espressions with respect to the scale factor $a$, we eventually get that the time behaviour of the BH dark energy is defined by the following differential equation:
\begin{widetext}
\begin{equation}\label{eq:BH_equation}
a \frac{d\,\Omega_{BH}(a)}{d\,a}= \Omega_{BH}(a) \left[ 1 - \Omega_{BH}(a) \right] \left\{ \left( 1+ \frac{\Delta}{2} \right) \mathcal{F}_{r}(a) +
\left( 1+ \Delta \right) \mathcal{F}_{m}(a) +
\left[1-\Omega_{BH}(a)\right]^{\frac{\Delta/2}{2\left(\frac{\Delta}{2}-1\right)}} \Omega_{BH}(a)^{\frac{1}{2\left(1-\frac{\Delta}{2} \right)}}  \mathcal{Q}(a) \right\}
\end{equation}
\end{widetext}
\begin{eqnarray}
\mathcal{F}_{r}(a) &=& \frac{2 \Omega_r a^{-4}}{\Omega_m a^{-3} + \Omega_r a^{-4}} \\
\mathcal{F}_{m}(a) &=& \frac{\Omega_m a^{-3}}{\Omega_m a^{-3} + \Omega_r a^{-4}} \nonumber \\
\mathcal{Q}(a) &=& 2 \left(1-\frac{\Delta}{2}\right) \left(H_0 \sqrt{\Omega_m a^{-3} + \Omega_r a^{-4}} \right)^{\frac{\Delta/2}{1-\frac{\Delta}{2}}} C^{\frac{1}{\frac{\Delta}{2}-1}} \nonumber
\end{eqnarray}

We need to note that (\ref{eq:BH_equation}) is different from (13) in our previous work \cite{Dabrowski:2020atl}, more specifically for the role and the expression of the radiation term. 

Thus, in this work we will perform two analysis. First, we will improve the results obtained in \cite{Dabrowski:2020atl} using the same data sets appearing there but employing the correct and updated version of BH DE which can be derived from the proper (\ref{eq:BH_equation}). Secondly, we will perform a new and fully comprehensive analysis using both the most updated cosmological data related to the cosmological background and the dynamical ones.

\subsection{Dynamical quantities}

While the application of BH DE to the cosmological background is quite straightforward, as we only need to rely on (\ref{eq:Hubble_alt}) inserting in it the solution of (\ref{eq:BH_equation}), the use of perturbation equations deserves a bit more of discussion.

The linear perturbation theory for Friedmann-Lema\^itre-Robertson-Walker universe was introduced by \cite{Lifshitz:1945du} and later summarized in several books and publications among which the seminal work of \cite{MUKHANOV1992203}. In order to derive the corresponding equations for DM perturbations within models with a DE component, whether a cosmological constant like $\Lambda$CDM or a dynamical DE fluid, one follows the standard procedure, i.e. to consider perturbed line elements of an expanding universe, which allow to get first order perturbed Einstein equations. Following \cite{1995ApJ...455....7M} one is then able to derive the first order energy-momentum conservation equations for a generic fluid with its equation of state parameter $w(a)$. In general one can derive the set of equations governing the growth of both the DM and DE perturbations as is summarized in (9-10) and (16-18) of \cite{PhysRevD.97.083522}. In the present work we follow the same procedure in the limit of no perturbations in DE. The resulting equations are scale invariant, are valid also for a $\Lambda$CDM cosmology, and read as \begin{eqnarray}
&&a^2\mathcal{H}^2\phi^{\prime \prime}+(4a\mathcal{H}^2+a\mathcal{\dot{H}})\phi^{\prime}+(\mathcal{H}^2+2\mathcal{\dot{H}})\phi=0, \nonumber \\
&&\delta_{m}^{\prime \prime}+ \frac{1}{a}\left(2+\frac{\mathcal{\dot{H}}}{a\mathcal{H}^2}\right)\delta_{m}^{\prime}=\frac{3\mathcal{H}^2}{2}\Omega_{ m}(a)\delta_{m},\label{eq:second-order-delta_m}
\end{eqnarray}
where: $\mathcal{H}=aH$ is the conformal Hubble parameter; $\delta_m$ is the density contrast parameter for DM; $\phi$ is the Bardeen potential coming from metric perturbation; the prime denotes derivative with respect to the scale factor; and the dot stands for the derivative with respect to cosmic time.
For the choice of the initial conditions we follow \cite{PhysRevD.79.023516}: for the matter density contrast and its derivative respectively, we use
\begin{eqnarray}
    \delta_{m}(a_i)&=&-2\phi(a_i)\left(1+\frac{1}{3\mathcal{H}(a_i)^2}\right ), \\
    \frac{ d \delta_{m}(a_i)}{ d a}&=&-\frac{2}{3}\frac{\phi(a_i)}{\mathcal{H}(a_i)^2}.
\end{eqnarray}
Similarly, as in \cite{10.1093/mnras/stv1478}, for the scalar field $\phi$ we set $\phi(a_i)=-6\times 10^{-7}$ and we assume $\phi^{\prime}(a_i)=0$, with $a_i=0.01$. Our results are unaffected by reasonable changes of the exact initial value of the $\phi$. 

The information about the growth rate of (matter) perturbations is then encoded in the quantity
\begin{align}\label{eq:growth_rate}
f(a) &= \frac{d\, \ln \delta_{m}(a)}{d\, \ln a}\, ,
\end{align}
derived from solving the set of differential equations ~(\ref{eq:second-order-delta_m}). Actually, observations from galaxy clustering are able to measure the combination $f\sigma_{8}(a) = f(a) \cdot \sigma_{8}(a)$, where $\sigma_{8}(a)$ is the conventionally defined amplitude of the linear power spectrum on the scale of $8\,h^{-1}$ Mpc:
\begin{align}\label{eq:sigma8}
\sigma_{8}(a) &= \sigma_{8,0}\frac{\delta_{m}(a)}{\delta_{m}(1)} \, ,
\end{align}
being $\sigma_{8,0}$ the normalization factor at present time $(a=1$ or equivalently $z=0)$.

\section{Constraining Barrow nonextensive holography by both kinematic and dynamical data}
\label{sec:Data}

For our statistical analysis we will consider many different type of cosmological probes, which we organize in different combinations. A quick overview is provided in Table~\ref{tab:data}.

One collection of data sets will be dubbed as ``late'', because it will not involve any calculation of physical quantities which are generally connected to early-times physics, like the sound horizon. Another case will be referred to as ``full'' because it will rely on the usage of both late-times and early-times data. Moreover, we perform analysis using only geometrical data (i.e. connected to the cosmological background), and in combination with dynamical data, which will take into account the growth of matter perturbations.

For what concerns the analysis which corrects the results previously published in \cite{Dabrowski:2020atl}, we have used the same (older) data for a consistent comparison. The only data in common with the present work are: the Cosmic Chronometers (CC), the Gamma Ray Bursts (GRBs) and the Baryon Acoustic Oscillations (BAO) data from the WiggleZ survey. Here we will describe in detail only the data which have been used for the analysis which we consider ``new and fully updated''. We thus refer the reader to the corresponding data section of \cite{Dabrowski:2020atl} for more details on the remaining data  used in that case.

\subsection{Pantheon+ SNeIa}

The most updated Type Ia Supernovae (SNeIa) data collection is the Pantheon+ sample \cite{Scolnic:2021amr,Peterson:2021hel,Carr:2021lcj,Brout:2022vxf}, made of $1701$ objects in the redshift range $0.001<z<2.26$. The $\chi^2_{SN}$ will be defined as
\begin{equation}\label{eq:chi_sn}
\chi^2_{SN} = \Delta \boldsymbol{\mathcal{\mu}}^{SN} \; \cdot \; \mathbf{C}^{-1}_{SN} \; \cdot \; \Delta  \boldsymbol{\mathcal{\mu}}^{SN} \;,
\end{equation}
where $\Delta\boldsymbol{\mathcal{\mu}} = \mathcal{\mu}_{\rm theo} - \mathcal{\mu}_{\rm obs}$ is the difference between the theoretical and the observed value of the distance modulus for each SNeIa and $\mathbf{C}_{SN}$ is the total (statistical plus systematic) covariance matrix. The distance modulus calculated from the theoretical model is:
\begin{equation}\label{mu_theo}
\mu_{theo}(z_{hel},z_{HD},\boldsymbol{p}) = 25 + 5 \log_{10} [ d_{L}(z_{hel}, z_{HD}, \boldsymbol{p}) ]\; ,
\end{equation}
where $d_L$ is the luminosity distance (in Mpc)
\begin{equation}
d_L(z_{hel}, z_{HD},\boldsymbol{p})=(1+z_{hel})\int_{0}^{z_{HD}}\frac{c\,dz'}{H(z',\boldsymbol{p})} \,,
\end{equation}
with: $H(z)$ the Hubble parameter (cosmological model dependent); $c$ the speed of light; $z_{hel}$ the heliocentric redshift; $z_{HD}$ the Hubble diagram redshift (i.e. the cosmic microwave background (CMB) redshift including peculiar velocity corrections \cite{Carr:2021lcj}); and $\boldsymbol{p}$ is the vector of cosmological parameters.

On the other hand, the observed distance modulus $\mu_{obs}$ is
\begin{equation}\label{mu_obs}
\mu_{obs} = m_{B} - \mathcal{M}\; ,
\end{equation}
with $m_{B}$ the standardized SNeIa blue apparent magnitude and $\mathcal{M}$ is the fiducial absolute magnitude calibrated by using primary distance anchors such as Cepheids. It is well known that in general $H_0$ and $\mathcal{M}$ are degenerate when SNeIa alone are used. But the Pantheon+ sample includes $77$ SNeIa located in galactic hosts for which the distance moduli can be measured from primary anchors (Cepheids), which means that the degeneracy can be broken and $H_0$ and $\mathcal{M}$ can be constrained separately. Thus, the vector $\Delta\boldsymbol{\mathcal{\mu}}$ will be
\begin{equation}
\Delta\boldsymbol{\mathcal{\mu}} = \left\{
  \begin{array}{ll}
    m_{B,i} - \mathcal{M} - \mu_{Ceph,i} & \hbox{$i \in$ Cepheid hosts} \\
    m_{B,i} - \mathcal{M} - \mu_{theo,i} & \hbox{otherwise,}
  \end{array}
\right.
\end{equation}
with $\mu_{Ceph}$ being the Cepheid calibrated host-galaxy distance
provided by the Pantheon+ team.

{\renewcommand{\tabcolsep}{1.5mm}
{\renewcommand{\arraystretch}{1.75}
\begin{table*}
\begin{minipage}{\textwidth}
\caption{Data sets used for the statistical analysis. For each probe we provide the name (first column) and the reference (last column) from which they are taken. The tick ``$\checkmark$'' means that the given data set is included in the final $\chi^2$ function. \textit{``Geo''} stands for \textit{``geometrical''} and only relates to the cosmological background; \textit{``+dyn''} includes solution of the perturbations equations. The term \textit{``late''} means that probes involving the calculation of physical quantities at recombination (and earlier) times are not included. The \textit{``full''} tag, instead, refers to the use of all the possible probes, also those ones connected to recombination (and earlier) times.} \label{tab:data}
\centering
\resizebox*{\textwidth}{!}{
\begin{tabular}{c|cc|cc|c}
\hline
 name  & geo-late & geo-full & geo-late+dyn & geo-full+dyn & ref. \\
\hline
Pantheon SNeIa   & \checkmark & \checkmark & \checkmark & \checkmark & \citet{Brout:2022vxf}\\
Cosmic Chronometers    & \checkmark & \checkmark & \checkmark & \checkmark & \citet{Jiao:2022aep}\\
GRBs   & \checkmark & \checkmark & \checkmark & \checkmark & \citet{Liu:2014vda}\\
CMB & $-$ & \checkmark & $-$ & \checkmark & \citet{Zhai:2019nad}  \\
SDSS-IV DR16 ELG & $-$ &  \checkmark (BAO) & \checkmark (RSD) & \checkmark (BAO+RSD) & \citet{Tamone:2020qrl,deMattia:2020fkb}\\
SDSS-III DR12 LRG & $-$ &  \checkmark (BAO) & \checkmark (RSD) & \checkmark (BAO+RSD) & \citet{BOSS:2016wmc}\\
SDSS-IV DR16 LRG & $-$ &  \checkmark (BAO) & \checkmark (RSD) & \checkmark (BAO+RSD) & \citet{Gil-Marin:2020bct,Bautista:2020ahg}\\
SDSS-IV DR16 LRG+Void & $-$ &  \checkmark (BAO) & \checkmark (RSD) & \checkmark (BAO+RSD) & \citet{Nadathur:2020vld}\\
SDSS-IV DR16 Lyman $\alpha$ & $-$ &  \checkmark (BAO) & $-$ & \checkmark (BAO) & \citet{duMasdesBourboux:2020pck}\\
SDSS-IV DR16 QSO (BAO) & $-$ & \checkmark (BAO) & \checkmark (RSD) & \checkmark (BAO+RSD) & \citet{Hou:2020rse,Neveux:2020voa}\\
SDSS-IV DR14 QSO (BAO) & $-$ & \checkmark (BAO) & \checkmark (RSD) & \checkmark (BAO+RSD) & \citet{Zhao:2018gvb}\\
WiggleZ & \checkmark (BAO) &  \checkmark (BAO) & \checkmark (BAO+RSD) & \checkmark (BAO+RSD) & \citet{Blake:2012pj}\\
\hline
2dFGRS  & $-$ & $-$ & \checkmark (RSD) & \checkmark (RSD) & \citet{Song:2008qt} \\
6dFGS    & $-$ & $-$ & \checkmark (RSD) & \checkmark (RSD) & \citet{Achitouv:2016mbn}\\
6dFGS Voids  & $-$ & $-$ & \checkmark (RSD) & \checkmark (RSD) & \citet{Achitouv:2016mbn}\\
FASTSOUND  & $-$ & $-$ & \checkmark (RSD) & \checkmark (RSD) & \citet{Okumura:2015lvp}\\
GAMA  & $-$ & $-$ & \checkmark (RSD) & \checkmark (RSD) & \citet{Blake:2013nif}\\
BOSS-WiggleZ  & $-$ & $-$ & \checkmark (RSD) & \checkmark (RSD) & \citet{Marin:2015ula} \\
BOSS LOWZ  & $-$ & $-$ & \checkmark (RSD) & \checkmark (RSD) & \citet{Lange:2021zre}\\
SDSS-IV DR15 LGR-SMALL & $-$ & $-$ & \checkmark (RSD) & \checkmark (RSD) & \citet{Chapman:2021hqe} \\
SDSS DR7 MGS & $-$ & $-$ & \checkmark (RSD) & \checkmark (RSD) & \citet{Howlett:2014opa}\\
VIPERS Voids  & $-$ & $-$ & \checkmark (RSD) & \checkmark (RSD) & \citet{Hawken:2016qcy} \\
VIPERS  & $-$ & $-$ & \checkmark (RSD) & \checkmark (RSD) & \citet{Mohammad:2018mdy}\\
VIPERS+GGL  & $-$ & $-$ & \checkmark (RSD) & \checkmark (RSD) & \citet{Jullo:2019lgq} \\
\hline
\hline
\end{tabular}}
\end{minipage}
\end{table*}}}

\subsection{Cosmic Chronometers}

As extensively outlined in \cite{Jimenez:2001gg,Moresco:2010wh,Moresco:2018xdr,Moresco:2020fbm,Moresco:2022phi}, early-type galaxies which undergo passive evolution and exhibit characteristic features in their spectra, can be defined and assessed as ``clocks'' or CC, and can provide measurements of the Hubble parameter $H(z)$ \cite{Moresco:2012by,Moresco:2012jh,Moresco:2015cya,Moresco:2016nqq,Moresco:2017hwt,Jimenez:2019onw,Jiao:2022aep}. The most updated sample of CC is from \cite{Jiao:2022aep} and spans the redshift range $0<z<1.965$. The corresponding $\chi^2_{H}$ can be written as
\begin{equation}\label{eq:chi_cc}
\chi^2_{H} = \Delta \boldsymbol{\mathcal{H}} \; \cdot \; \mathbf{C}^{-1}_{H} \; \cdot \; \Delta  \boldsymbol{\mathcal{H}} \;,
\end{equation}
where $\Delta \boldsymbol{\mathcal{H}} = H_{theo} - H_{data}$ is the difference between the theoretical and observed Hubble parameter, and $\mathbf{C}_{H}$ is the total (statistical plus systematics) covariance matrix calculated following prescriptions from \cite{Moresco:2020fbm}.

\subsection{Gamma Ray Bursts}

The so-called ``Mayflower'' sample \cite{Liu:2014vda}, overcomes the well-known issue of calibration of GRBs by relying on a robust cosmological model independent procedure. It is made of 79 GRBs in the redshift interval $1.44<z<8.1$ for which we recover the distance modulus. The $\chi_{G}^2$ is defined exactly like in the SNeIa case, (\ref{eq:chi_sn}), but in this case we cannot disentangle between the Hubble constant and the absolute magnitude, so that we have to marginalize over them. Following \cite{Conley:2011ku} it becomes
\begin{equation}\label{eq:chi_grb}
\chi^2_{GRB}=a+\log d/(2\pi)-b^2/d\,,
\end{equation} 
with $a\equiv \left(\Delta\boldsymbol{\mathcal{\mu}}_{G}\right)^T \, \cdot \, \mathbf{C}^{-1}_{G} \, \cdot \, \Delta  \boldsymbol{\mathcal{\mu}}_{G}$, $b\equiv\left(\Delta \boldsymbol{\mathcal{\mu}}_{G}\right)^T \, \cdot \, \mathbf{C}^{-1}_{G} \, \cdot \, \boldsymbol{1}$ and $d\equiv\boldsymbol{1}\, \cdot \, \mathbf{C}^{-1}_{G} \, \cdot \, \boldsymbol{1}$.

\subsection{Cosmic Microwave Background}

The Cosmic Microwave Background (CMB) analysis is not performed using the full power spectra provided by \textit{Planck} \cite{Planck:2018vyg} but instead using the shift parameters defined in \cite{Wang:2007mza} and derived from the latest \textit{Planck} $2018$ data release in \cite{Zhai:2019nad}. The $\chi^2_{CMB}$ is defined as
\begin{equation}
\chi^2_{CMB} = \Delta \boldsymbol{\mathcal{F}}^{CMB} \; \cdot \; \mathbf{C}^{-1}_{CMB} \; \cdot \; \Delta  \boldsymbol{\mathcal{F}}^{CMB} \; ,
\end{equation}
where the vector $\mathcal{F}^{CMB}$ corresponds to the quantities:
\begin{eqnarray} \label{eq:CMB_shift}
R(\boldsymbol{p}) &\equiv& \sqrt{\Omega_m H^2_{0}} \frac{r(z_{\ast},\boldsymbol{p})}{c}, \nonumber \\
l_{a}(\boldsymbol{p}) &\equiv& \pi \frac{r(z_{\ast},\boldsymbol{p})}{r_{s}(z_{\ast},\boldsymbol{p})}\,,
\end{eqnarray}
in addition to a constraint on the baryonic content, $\Omega_b\,h^2$, and on the dark matter content, $(\Omega_m-\Omega_b)h^2$. In (\ref{eq:CMB_shift}), $r_{s}(z_{\ast},\boldsymbol{p})$ is the comoving sound horizon evaluated at the photon-decoupling redshift. The general definition of the comoving sound horizon is
\begin{equation}\label{eq:soundhor}
r_{s}(z,\boldsymbol{p}) = \int^{\infty}_{z} \frac{c_{s}(z')}{H(z',\boldsymbol{p})} \mathrm{d}z'\, ,
\end{equation}
where the sound speed is given by
\begin{equation}\label{eq:soundspeed}
c_{s}(z) = \frac{c}{\sqrt{3(1+\overline{R}_{b}\, (1+z)^{-1})}} \; ,
\end{equation}
with the baryon-to-photon density ratio parameters defined as $\overline{R}_{b}= 31500 \Omega_{b} \, h^{2} \left( T_{CMB}/ 2.7 \right)^{-4}$ and $T_{CMB} = 2.726$ K. The photon-decouping redshift is evaluated using the fitting formula from \cite{Hu:1995en},
\begin{eqnarray}{\label{eq:zdecoupl}}
z_{\ast} &=& 1048 \left[ 1 + 0.00124 (\Omega_{b} h^{2})^{-0.738}\right] \times   \nonumber \\
&& \left(1+g_{1} (\Omega_{m} h^{2})^{g_{2}} \right)  \,,
\end{eqnarray}
where the factors $g_1$ and $g_2$ are given by
\begin{eqnarray}
g_{1} &=& \frac{0.0783 (\Omega_{b} h^{2})^{-0.238}}{1+39.5(\Omega_{b} h^{2})^{-0.763}}\,, \nonumber \\
g_{2} &=& \frac{0.560}{1+21.1(\Omega_{b} h^{2})^{1.81}} \,.
\end{eqnarray}

Finally, $r(z_{\ast}, \boldsymbol{p})$ is the comoving distance at decoupling, i.e. using the definition of the comoving distance:
\begin{equation}\label{eq:comovdist}
d_{M}(z,\boldsymbol{p})=\int_{0}^{z} \frac{c\, dz'}{H(z',\boldsymbol{p})} \; ,
\end{equation}
we set $r(z_{\ast},\boldsymbol{p}) = d_M(z_{\ast},\boldsymbol{p})$.

\subsection{Baryon Acoustic Oscillations} 

For BAO we consider multiple data sets from different surveys. In general, the $\chi^2$ is defined as
\begin{equation}
\chi^2_{BAO} = \Delta \boldsymbol{\mathcal{F}}^{BAO} \, \cdot \ \mathbf{C}^{-1}_{BAO} \, \cdot \, \Delta  \boldsymbol{\mathcal{F}}^{BAO} \ ,
\end{equation}
with the observables $\mathcal{F}^{BAO}$ which change from survey to survey.

When we employ the data from the WiggleZ Dark Energy Survey \cite{Blake:2012pj}, at redshifts $z=\{0.44, 0.6, 0.73\}$, the relevant physical quantities are the acoustic parameter
\begin{equation}\label{eq:AWiggle}
A(z,\boldsymbol{p}) = 100  \sqrt{\Omega_{m} \, h^2} \frac{d_{V}(z,\boldsymbol{p})}{c \, z} \, ,
\end{equation}
where $h=H_0/100$, and the Alcock-Paczynski distortion parameter
\begin{equation}\label{eq:FWiggle}
F(z,\boldsymbol{p}) = (1+z)  \frac{d_{A}(z,\boldsymbol{p})\, H(z,\boldsymbol{p})}{c} \, ,
\end{equation}
where $d_{A}$ is the angular diameter distance defined as
\begin{equation} \label{eq:ang_dist}
d_{A}(z,\boldsymbol{p})=\frac{1}{1+z}\int_{0}^{z} \frac{c\, dz'}{H(z',\boldsymbol{p})} \;,
\end{equation}
and
\begin{equation}
d_{V}(z,\boldsymbol{\theta})=\left[ (1+z)^2 d^{2}_{A}(z,\boldsymbol{\theta}) \frac{c z}{H(z,\boldsymbol{\theta})}\right]^{1/3}
\end{equation}
is the geometric mean of the radial and tangential BAO modes. This data set is independent of any early-times quantity an it is thus the only BAO data set included in the late-times analysis.

We also consider data from multiple analysis of the latest release of the Sloan Digital Sky Survey (SDSS) Extended Baryon Oscillation Spectroscopic Survey (eBOSS) observations. Each one of the following data are used for the full data analysis and not for the late-times one because they involve the calculation of the sound horizon at early-times.

For all the SDSS data \cite{Tamone:2020qrl,deMattia:2020fkb,BOSS:2016wmc,Gil-Marin:2020bct,Bautista:2020ahg,Nadathur:2020vld,duMasdesBourboux:2020pck,Hou:2020rse,Neveux:2020voa}, with the exception of the SDSS-IV DR14 quasars analysis from \cite{Zhao:2018gvb}, the following quantities are given:
\begin{equation}
\frac{d_{M}(z,\boldsymbol{p})}{r_{s}(z_{d},\boldsymbol{p})}, \qquad \frac{c}{H(z,\boldsymbol{p}) r_{s}(z_{d},\boldsymbol{p})} \,,
\end{equation}
where the comoving distance $d_M$ is given by (\ref{eq:comovdist}) and the sound horizon is evaluated at the dragging redshift $z_{d}$. The dragging redshift is estimated using the analytical approximation provided in  \cite{Eisenstein:1997ik} which reads
\begin{equation}\label{eq:zdrag}
z_{d} = \frac{1291 (\Omega_{m} \, h^2)^{0.251}}{1+0.659(\Omega_{m} \, h^2)^{0.828}} \left[ 1+ b_{1} (\Omega_{b} \, h^2)^{b2}\right]\; ,
\end{equation}
where the factors $b_1$ and $b_2$ are given by
\begin{eqnarray}\label{eq:zdrag_b}
b_{1} &=& 0.313 (\Omega_{m} \, h^2)^{-0.419} \left[ 1+0.607 (\Omega_{m} \, h^2)^{0.6748}\right] \,,
	\nonumber \\
b_{2} &=& 0.238 (\Omega_{m} \, h^2)^{0.223}\,.
\end{eqnarray}

Data from \cite{Zhao:2018gvb} are instead expressed in terms of 
\begin{equation}
d_{A}(z,\boldsymbol{p}) \frac{r^{fid}_{s}(z_{d},)}{r_{s}(z_{d},\boldsymbol{p})}, \qquad H(z,\boldsymbol{p}) \frac{r_{s}(z_{d},\boldsymbol{p})}{r^{fid}_{s}(z_{d},\boldsymbol{p})} \,,
\end{equation}
where $r^{fid}_{s}(z_{d})$ is the sound horizon at dragging redshift calculated for the given fiducial cosmological model considered in \cite{Zhao:2018gvb}, which is equal to $147.78$ Mpc.

\subsection{Redshift Space Distorsions}

For the dynamical analysis based on the perturbation equations and thus connected to redshift space distorsions (RSD) measurements, we have used the data sets (with the \textit{RSD} tag) which we only report in Table~\ref{tab:data}, for the sake of legibility and clarity. Note that although in some literature \cite{Skara:2019usd,Benisty:2020kdt} larger samples are considered, we have decided to retain only those data which are unequivocally unrelated and to discard those which are superseded by most updated ones, as it happens in the case of the final release from SDSS.
       
One important comment is in order here: RSD measurement are not cosmologically independent, which means that data points are provided for a given fiducial cosmology. Thus, in order to be used by us in our analysis for our Barrow DE model, we have to re-scale them, considering the following relation:
$$
\left[f\sigma_{8}(z)\right]_{model} = \left[f\sigma_{8}(z)\right]_{data} \frac{H_{fid,\,data}(z)\cdot D_{A/fid,\,data}(z)}{H_{model}(z)\cdot D_{A/ model}(z)}\,
$$
where $D_{A}$ is the angular diameter distance defined in (\ref{eq:ang_dist}).

\subsection{Statistical tools}

The total $\chi^2$ properly corresponding to each data combination is minimized using our own code for Monte Carlo Markov Chain (MCMC). The convergence of the chains is checked using the diagnostic described in \citep{Dunkley:2004sv}. 

In order to establish the reliability of the Barrow DE with respect to the standard $\Lambda$CDM scenario, we calculate the Bayes Factor \cite{doi:10.1080/01621459.1995.10476572}, $\mathcal{B}^{i}_{j}$, defined as the ratio between the Bayesian Evidences of the two compared models, in our case model $\mathcal{M}_i$ being the Barrow one, and model $\mathcal{M}_j$ being the $\Lambda$CDM. We calculate the evidence numerically using our own code implementing the Nested Sampling algorithm developed by \cite{Mukherjee:2005wg}. Finally, the interpretation of the Bayes Factor is conducted using the empirical Jeffrey's scale \cite{Jeffreys1939-JEFTOP-5}. 

\section{Discussion and Conclusions}
\label{sec:Results}

In this work we have applied a complete set of cosmological data -- both kinematic and dynamical -- to the Barrow entropy holographic screen. There are a couple of interesting results which we list below. 

Preliminary, we would briefly comment how the application of the new data influences the results of our previous paper \cite{Dabrowski:2020atl} on the Barrow entropy. In fact, comparing Table~\ref{tab:results} here with Table~I of \cite{Dabrowski:2020atl}, we can clearly see how the corrected (\ref{eq:BH_equation}) implies no change at all for what concerns the analysis with the late-times data. Both results are fully compatible and totally statistically consistent. Instead, the greatest differences appear when we use full data. As they are totally in line with the results which we get from the newest and most updated data set, we focus on a general discussion about them in the next paragraphs.

First of all, all data tests lead to the conclusion that the Barrow fractal index $\Delta$ is {\it bound from below}, being $\Delta > 0.86$ for full data and peaking towards $\Delta \to 1$ in all our data combination, which means that cosmological horizon should be of the fractal nature. This has a very interesting consequence in view of the nonextensive entropies cosmological applications -- namely, the better fit to the data, the more extensivity in the system. Otherwise, we might state that the resolution of the dark energy problem due to the Barrow (or Tsallis-Cirto) holography is possible if the thermodynamics of the universe is more like the classical Gibbs one, which is extensive and additive. However, unlike Gibbs, Barrow entropy is still nonadditive.

If looked at from the perspective of a cosmologist, we might say that this result does not come totally unforeseen. Indeed, we know very well that the $\Lambda$CDM model, i.e. a cosmological constant for what concerns the dark energy component, provides the most successful fit to the cosmological data we have considered. Thus, a \textit{biased implicit preference} toward a cosmological constant dynamical behaviour might be expected, although not necessarily obvious. Considering that the Barrow DE, as stated in previous sections, can achieve a cosmological constant behaviour \textit{only} at $\Delta=2$, which lies beyond the expected physical range for this scenario, the peak toward the maximum upper limit $\Delta=1$ can be seen as an \textit{implicit trend} from the data towards a cosmological constant behaviour.

Another very important point to stress is that the ``standard'' limit of a non-fractal horizon, i.e. $\Delta = 0$, is excluded by our set of data. In fact, our bound on Barrow index $\Delta = 2(\delta - 1)$ contradicts other evaluations from literature.

The difference in the retrieved estimations is very likely related to the different approaches which are used to include entropy definitions in the cosmological context. More specifically, one can identify two main ways: by the holographic principle, which is also used in this work, and by the gravity-thermodynamics conjecture \cite{Jacobson:1995ab,Padmanabhan:2003gd,Cai:2005ra,Padmanabhan:2009vy}.

In the gravity-thermodynamics-based works, the new contributions to the first Friedmann equation due to new definition of entropy can be written as two terms: a function of scale factor (which is mostly a function of the Hubble parameter $H(a)$) and a constant (of integration). The latter is always identified with the cosmological constant. And we know very well that with a cosmological constant one can describe quite well and in its entirety the dark energy sector and its dynamics. From this point of view, entropic contributions, at the cosmological level, are almost like a ``nuisance''. Furthermore, in gravity-thermodynamics analysis, the authors are generally interested into an epoch where dark energy is negligible and they mostly assume that the only contribution to the energy-matter is from radiation. Thus, what it is really going to be tested with the entropic contributions are ``corrections to radiation physics'', which one could naively expect to be small.

Given such premises, in \cite{Luciano:2022hhy}, the constraints on Barrow entropy are derived from observational bounds on baryon asymmetry, leading to $\Delta \sim 0.005-0.008$. In \cite{Luciano:2022ely} the same analysis is performed for Tsallis entropy, with a final constraint of $0.002 \lesssim |\delta-1| \lesssim 0.004$. In \cite{Ghoshal:2021ief}, Tsallis entropy is confronted with Big Bang nucleosynthesis, and it is found that $1-\delta<10^{-5}$. The same data are confronted with Barrow entropy in \cite{Barrow:2020kug}, and the authors get $\Delta \lesssim 1.4 \cdot 10^{-4}$.

In \cite{Leon:2021wyx} the entropic modified Friedmann equation within the gravity-thermodynamics approach is confronted with a set of cosmological probes, including Pantheon SNeIa and a BAO sample, but the early-times physics (and thus the role of radiation) is not included in their model, differently from our (\ref{eq:BH_equation}). After applying a gaussian prior on the sound horizon, on $\Omega_m$ and $H_0$, they obtain $\Delta \sim 10^{-4}$. In \cite{Asghari:2021lzu}, the authors start from the gravity-thermodynamics approach again, and although they perform a more detailed analysis of the consequence of Tsallis entropy at the level of cosmological perturbations, they also introduce a scalar field with constant equation of state which behaves as a dark energy fluid. Thus, entropic corrections result to bring a negligible contribution in the cosmological context, with the fluid having a $w \approx -1$ and the entropic parameter $\delta\approx 0.9997$. Finally, confronting Barrow entropy with inflationary cosmological parameters constraint from \textit{Planck}, in \cite{Luciano:2023roh} the Barrow parameter is qualitatively set as $\Delta \lesssim 10^{-4}$ (but assuming a number of e-fold of $30$).

When it comes to the application of the holographic principle, as we have done in this work, results are still different from what we find. In \cite{FeiziMangoudehi:2022rwj} the authors do not use any early-times probe, but only SNeIa, CC and GRBs, and a different horizon from ours, and their final estimation for the Tsallis parameter is $\delta \approx 0.16$, which corresponds to $\Delta \approx -1.68$, clearly out of the physical boundary required by Barrow theory. In \cite{Sadri:2019qxt} the holographic principle is applied also to our same horizon definition, and there is the usage of early-times data (BAO and CMB), but the contribution of radiation seems to be missing from the main equations derived for Barrow entropy. Anyway, they constrain $\delta \approx 1.07$ corresponding to $\Delta \approx 0.14$. Similar results are obtained in \cite{Saridakis:2018unr,Anagnostopoulos:2020ctz} using only late-times data, with $\Delta \sim 0.09$, but with the holographic parameter $C$ being totally unconstraine (see their Fig.~2). Finally, in \cite{Adhikary:2021xym}, Barrow holographic dark energy is compared to SNeIa and CC only, but assuming a varying spatial curvature; final ranges for Barrow entropy parameter are $\Delta \sim 0.06 \div 0.2$.

As for all the other cosmological parameters, we can note that they are always perfectly consistent with $\Lambda$CDM. A trend towards slightly lower values for $\Omega_m$ seems to be favoured by the Barrow scenario, but the statistical significance of this deviation is negligible.

We can also note how the holographic parameter $C$ is well restricted to a quite precise range, being $\sim5$ independently of the probes which are considered, with smaller errors when the full sample is used. The interpretation of such numbers, although, are not an easy task, because the holographic parameter actually has no defined units.

Finally, in general, if we look at the Bayes Factors we can state that the Barrow DE model is always statistically disfavoured with respect to $\Lambda$CDM, with the inclusion of early-times and dynamical probes leading to the most penalizing results. One interesting point to address would be if there were any positive contributions to the cosmological tensions which are now under scrutiny and debate. But we can easily see that there is no real improvement in that for $H_0$, while for $S_{8,0}= \sigma_{8,0} \sqrt{\Omega_m/0.3}$ we can see some mildly positive effects namely, even including \textit{Planck} CMB data, we get a value of $S_8$ which is consistent with late-times large scale surveys.

{\renewcommand{\tabcolsep}{1.5mm}
{\renewcommand{\arraystretch}{2.}
\begin{table*}
\begin{minipage}{\textwidth}
\caption{Results from the statistical analysis. For each parameter we provide the median and the $1\sigma$ constraints. The columns show: $1.$ considered theoretical scenario; $2.$ dimensionless matter parameter, $\Omega_m$; $3.$ dimensionless baryonic parameter, $\Omega_b$; $4.$ dimensionless Hubble constant, $h$; $5.$ fiducial absolute magnitude, $\mathcal{M}$; $6.$ amplitude of the linear power spectrum at present time, $\sigma_{8,0}$; $7.$ Barrow entropic parameter, $\Delta$; $8.$ holographic parameter, $C$; $9.$ amplitude of the weak lensing measurement (secondary derived parameter), $S_{8,0}$; $10.$ logarithm of the Bayes Factor, $\log \mathcal{B}^{i}_{j}$.}\label{tab:results}
\centering
\resizebox*{\textwidth}{!}{
\begin{tabular}{c|ccccccc|cc}
\hline
   & $\Omega_m$ & $\Omega_b$ & $h$ & $\mathcal{M}$ & $\sigma_{8,0}$ & $\Delta$ & $C$ & $S_{8,0}$ & $\log \mathcal{B}^{i}_{j}$ \\
\hline
\hline
\multicolumn{10}{c}{``Revision'' of \cite{Dabrowski:2020atl}} \\
\hline 
LCDM (geo-late) & $0.293^{+0.016}_{-0.016}$ & $-$ & $0.713^{+0.013}_{-0.013}$ & $-$ & $-$ & $-$ & $-$ & $-$ & $\mathit{0}$ \\
LCDM (geo-full) & $0.319^{+0.005}_{-0.005}$ & $0.0494^{+0.0004}_{-0.0004}$ & $0.673^{+0.003}_{-0.003}$ & $-$ & $-$ & $-$ & $-$ & $-$ & $\mathit{0}$ \\
\hline
BH1 (geo-late) & $0.290^{+0.020}_{-0.019}$ & $-$ & $0.715^{+0.014}_{-0.013}$ & $-$ & $-$ & $>0.63$ & $3.93^{+1.77}_{-1.88}$ & $-$ & $-0.71^{+0.03}_{-0.02}$ \\
BH1 (geo-full) & $0.314^{+0.006}_{-0.006}$ & $0.049^{+0.001}_{-0.001}$ & $0.676^{+0.007}_{-0.007}$ & $-$ & $-$ & $>0.84$ & $4.66^{+0.87}_{-1.07}$ & $-$ & $-0.05^{+0.03}_{-0.03}$ \\
\hline
\hline
\multicolumn{10}{c}{Updated and newest constraints} \\
\hline
LCDM (geo-late)   & $0.321^{+0.015}_{-0.015}$ & $-$ & $0.730^{+0.010}_{-0.009}$ & $-19.263^{+0.028}_{-0.028}$ & $-$ & $-$ & $-$ & $-$ & $\mathit{0}$ \\
LCDM (geo-full)   & $0.318^{+0.007}_{-0.006}$ & $0.0493^{+0.0006}_{-0.0006}$ & $0.674^{+0.004}_{-0.004}$ & $-19.437^{+0.012}_{-0.012}$ & $-$ & $-$ & $-$ & $-$ & $\mathit{0}$\\
LCDM (geo-late+dyn) & $0.315^{+0.014}_{-0.014}$ & $-$ & $0.731^{+0.010}_{-0.010}$ & $-19.263^{+0.028}_{-0.028}$ & $0.770^{+0.018}_{-0.017}$  & $-$ & $-$ & $0.790^{+0.023}_{-0.022}$ & $\mathit{0}$\\
LCDM (geo-full+dyn) & $0.314^{+0.006}_{-0.005}$ & $0.0490^{+0.0006}_{-0.0006}$ & $0.677^{+0.004}_{-0.004}$ & $-19.429^{+0.011}_{-0.011}$ & $0.779^{+0.017}_{-0.017}$ & $-$ & $-$ & $0.796^{+0.019}_{-0.019}$ & $\mathit{0}$ \\
\hline
BH (geo-late) & $0.300^{+0.020}_{-0.019}$ & $-$ & $0.729^{+0.010}_{-0.010}$ & $-19.263^{+0.028}_{-0.029}$ & $-$ & $>0.63$ & $4.50^{+2.20}_{-2.13}$ & $-$ & $-0.39^{+0.02}_{-0.04}$ \\
BH (geo-full) & $0.311^{+0.006}_{-0.006}$ & $0.0486^{+0.0008}_{-0.0008}$ & $0.679^{+0.006}_{-0.006}$ & $-19.438^{+0.013}_{-0.013}$ & $-$ & $>0.82$ & $4.58^{+0.90}_{-1.16}$ & $-$ & $-2.99^{+0.04}_{-0.04}$ \\
BH (geo-late+dyn) & $0.290^{+0.018}_{-0.017}$ & $-$ & $0.729^{+0.010}_{-0.010}$ & $-19.261^{+0.028}_{-0.028}$ & $0.791^{+0.022}_{-0.022}$ & $>0.69$ & $5.31^{+1.97}_{-2.27}$ & $0.791^{+0.022}_{-0.022}$ & $-0.35^{+0.03}_{-0.03}$\\
BH (geo-full+dyn) & $0.307^{+0.006}_{-0.006}$ & $0.0484^{+0.0009}_{-0.0008}$ & $0.681^{+0.006}_{-0.005}$ & $-19.431^{+0.013}_{-0.013}$ & $0.777^{+0.017}_{-0.017}$ & $>0.86$ & $4.89^{+0.76}_{-1.03}$ & $0.786^{+0.020}_{-0.020}$ & $-4.36^{+0.04}_{-0.04}$ \\
\end{tabular}}
\end{minipage}
\end{table*}}}

\bibliographystyle{apsrev4-1}
\bibliography{biblio}

\end{document}